\documentclass[letterpaper,10pt]{article}
\usepackage{authblk}
\usepackage{amsfonts}
\usepackage{fixmath}
\usepackage{setspace}
\usepackage{enumitem}
\usepackage{amsmath}
\usepackage{algorithmic}
\usepackage{algorithm,xcolor}

\newcounter{myprot}
\newcounter{myalg}
\newcounter{myexam}
\newenvironment{myexam}
{\refstepcounter{myexam}  \noindent \textbf{EXAMPLE \arabic{myexam}:}}
{\vspace{.25em}}

\newcounter{mythm}
\newenvironment{mythm}
{\refstepcounter{mythm}  \noindent \textbf{THEOREM \arabic{mythm}:}\em}
{\vspace{.25em}}

\newcounter{mylem}
\newenvironment{mylem}
{\refstepcounter{mylem}  \noindent \textbf{LEMMA \arabic{mylem}:}\em}
{\vspace{.25em}}

\newcounter{mycor}
\newenvironment{mycor}
{\refstepcounter{mycor} \vspace{1em} \noindent \textbf{COROLLARY \arabic{mycor}:}}
{\vspace{.5em}}

\newcounter{myobs}
\newcounter{mydef}
\newenvironment{mydef}
{\refstepcounter{mydef} \vspace{1em} \noindent \textbf{DEFINITION \arabic{mydef}:}}
{\vspace{.5em}}

\newcounter{myconj}
\newenvironment{myproof}
{\noindent \textbf{PROOF:}}
{\vspace{-3ex}\begin{flushright} $\Box$ \end{flushright}\vspace{2ex}}

\usepackage{mathtools,url}
\usepackage{latexsym}
\usepackage{subcaption} 
\usepackage[font={small,sf}]{caption}

\makeatletter
\g@addto@macro{\UrlBreaks}{\UrlOrds}
\makeatother

\renewcommand{\footnotesize}{\fontsize{8pt}{10pt}\selectfont}

\newif\ifinappendix
\let\oldappendix\appendix
\renewcommand{\appendix}{
  \oldappendix
  \inappendixtrue
}

\usepackage{booktabs}
\usepackage[hidelinks]{hyperref}
\usepackage{xfrac}
\usepackage{wrapfig,multirow,relsize}
\usepackage[nospace,compress]{cite}
\usepackage{microtype}

\newcommand{\OmitText}[1]{ {} }

\usepackage{xspace}

\newcommand{\1}{{\em (i)}}
\newcommand{\2}{{\em (ii)}}
\newcommand{\3}{{\em (iii)}}

\usepackage[all]{hypcap}

\usepackage[capitalise,nameinlink]{cleveref}
\crefname{section}{Sect.}{Sect.}
\Crefname{section}{Section}{Sections}

\usepackage{url}
\makeatletter
\g@addto@macro{\UrlBreaks}{\UrlOrds}
\makeatother
\makeatletter
\def\Url@twoslashes{\mathchar`\/\@ifnextchar/{\kern-.2em}{}}
\g@addto@macro\UrlSpecials{\do\/{\Url@twoslashes}}
\makeatother

\begin{document}

\title{\Large Greedy but Cautious: Conditions for Miner Convergence to Resource Allocation Equilibrium}

\author[1]{\small George Bissias}
\author[1]{\small Brian N. Levine}
\author[2]{\small David Thibodeau}

\affil[1]{\footnotesize College of Information and Computer Sciences,  UMass Amherst\\
\texttt{{\{gbiss,levine\}@cs.umass.edu}}}
\affil[2]{\footnotesize Florida Department of Corrections\\

\texttt{davidpthibodeau@gmail.com}}

\maketitle

\setlength{\belowdisplayskip}{3pt} 
\setlength{\belowdisplayshortskip}{3pt}
\setlength{\abovedisplayskip}{3pt} 
\setlength{\abovedisplayshortskip}{3pt}

\begin{abstract}
All public blockchains are secured by a proof of opportunity cost among block producers. For example, the security offered by proof-of-work (PoW) systems, like Bitcoin, is due to spent computation; it is work precisely because it cannot be performed for free.  In general,  more resources provably lost in producing blocks yields more security for the blockchain. When two blockchains share the same mechanism for providing opportunity cost, as is the case when they share the same PoW algorithm, the two chains compete for resources from block producers. Indeed, if there exists a liquid market between resource types, then theoretically all blockchains will compete for resources. In this paper, we show that there exists a resource allocation equilibrium between any two blockchains, which is essentially driven by the fiat value of reward that each chain offers in return for providing security. We go on to prove that this equilibrium is singular and always achieved provided that block producers behave in a greedy, but cautious fashion. The opposite is true when they are overly greedy: resource allocation oscillates in extremes between the two chains. We show that these results hold both in practice and in a block generation simulation. Finally, we demonstrate several applications of this theory including a trustless price-ratio oracle, increased security for blockchains whose coins have lower fiat value, and a quantification of cost to allocating resources away from the equilibrium.
\end{abstract}

\section{Introduction}

Cryptocurrencies such as Bitcoin~\cite{Nakamoto:2009} are a confluence of systems engineering, economics, and game theory. In some ways, cryptocurrency economics is a prosaic application of classic economic theory. For example, the economy defined by Bitcoin has an extremely simple monetary policy: a fixed coin issuance schedule, which makes inflation entirely predictable\footnote{Coin destruction, through loss of private keys, is much more difficult to measure.}. However, the procedural properties of cryptocurrencies, more software than policy, give rise to remarkably crisp economic tradeoffs that manifest surprisingly regular macro-level phenomena. In this paper, we examine one such phenomenon that arises from the dynamics of the so-called difficulty adjustment algorithms (DAAs) used by blockchains that employ proof-of-work (PoW) for security. We show that, as predicted by Spiegelman et al.~\cite{Spiegelman:2018}, but in contrast to the model of Kwon et al.~\cite{Kwon:2019}, much of aggregate miner behavior can be explained by their proclivity to increase immediate profit. More specifically, we offer a rationale for the division of hash rate that is manifest between two blockchains that share the same PoW algorithms. Remarkably, the same reasoning generalizes to the division of security resources between blockchains using different PoW algorithms and even using different consensus mechanisms altogether. We make the following contributions over prior work.

\begin{itemize}
\item We define a hash rate \emph{allocation equilibrium} between two blockchains that use the same proof-of-work (PoW) algorithm. We show that a unique allocation equilibrium exists and that it aligns with one of the Nash equilibria described by Kwon et al.~\cite{Kwon:2019}. The allocation equilibrium is much less stringent than a Nash equilibrium. The former assumes only that miners individually tend to act to maximize their profit according to a metric similar to the popular \emph{difficulty adjusted reward index} (DARI). In contrast, the Nash equilibrium assumes that miners have knowledge of a complex utility function and the strategies of other miners at equilibrium. 
\item We prove the conditions under which hash rate allocation will converge to the allocation equilibrium and anticipate allocation dynamics when these conditions are not met.
\item We show that, given an efficient market for buying and selling PoW hash rate, the allocation equilibrium generalizes to pairs of blockchains that use different PoW algorithms. We also show that the existence of efficient hash rate markets allows for generalization to equilibria between PoW and PoS blockchains.
\item We empirically validate the existence of allocation equilibria between several of the top blockchain projects that share the same PoW algorithm, including BTC versus BCH, and ETH versus ETC. Their adherence to the equilibrium is found to be quite strong. We also corroborate our theoretical results in simulation, showing precisely the conditions under which pairs of blockchains converge to, or diverge from, the allocation equilibrium.
\item We provide several applications for the allocation equilibrium in the real world, including: a trustless price-ratio oracle, increasing security for minority hash rate blockchains, and measuring the cost to miners who provide security to a blockchain beyond the equilibrium.
\end{itemize}

\section{Related Work}
\label{sec:related}

Prat and Walter~\cite{Prat:2018} modeled the impacts of mining difficulty and coin exchange rate on mining profitability for a single blockchain. They found that miners will not purchase new mining hardware if the fiat value of the coinbase reward is insufficient to accommodate the resulting rise in mining difficulty. The paper demonstrates empirically that this relationship holds quite well in practice for data ranging from 2012 until 2018. Furthermore, in the context of a single blockchain, Ma et al.~\cite{June:2019} showed that there exists a Nash equilibrium for the computing power allocated by miners given a fixed mining difficulty.

The work of Kwon et al.~\cite{Kwon:2019} is similar to ours. They showed that there exist multiple Nash equilibria for miners who allocate their hash rate among two bockchains sharing the same PoW and inter-block time. One equilibrium, $\mathcal{E}_{\text{econ}}$ (which coincides with our allocation equilibrium), exists at the relative price ratios of the two coins, but for the others, which we denote collectively as $\mathcal{E}_{\text{others}}$, no subset of economically rational miners will dedicate their hash rate to the \emph{discounted} blockchain having the lower coin price. And if a sufficiently large fraction of miners commit to mining the discounted blockchain, then they will be alone in mining on that chain. These results suggest the possibility that, in the presence of mostly rational miners, the discounted blockchain may end up supported solely by a centralized cabal of committed miners. Kwon et al.~\cite{Kwon:2019} further reported a tendency for profit-seeking miners to generally move toward $\mathcal{E}_{\text{econ}}$, but argued that they can also be pulled toward equilibria $\mathcal{E}_{\text{others}}$, and it is not clear how these dynamics play out in an iterative game. Moreover, their model involves a complex utility function, sophisticated strategies, and requires that each miner knows the strategy of the others in order to maintain equilibrium. In the present work, we assume only that a certain fraction of miners act so as to increase their immediate profit. We show formally that, under those conditions, hash rate allocated to the discounted chain will always converge to $\mathcal{E}_{\text{econ}}$. This result implies that the discounted chain can count on a minimum hash rate proportional to its coin's price relative to competing chains. Moreover, we demonstrate empirically and in simulation the conditions under which convergence succeeds. Our analysis also generalizes to equilibria between blockchains using different PoW algorithms and even those using proof-of-stake (PoS).

Also closely related is the work of Spiegelman et al.~\cite{Spiegelman:2018} who apply the theory of Potential Games~\cite{Monderer:1996} to the problem of miner hash rate allocation across multiple blockchains. They prove that, regardless of individual hash rate and coinbase rewards for each of the blockchains, hash rate allocation will converge to a pure equilibrium provided that miners follow \emph{better response learning}. The model assumes ``minimal rationality on behalf of the players, i.e., that they follow an arbitrary better response step improving their individual payoffs.'' Spiegelman et al.~\cite{Spiegelman:2018} do not identify a specific equilibrium point, nor do they specify what the \emph{better response} should be. But their work anticipates some of the theoretical results we present in Section~\ref{sec:dynamics}. Furthermore, they show that the equilibrium point can be changed by changing a blockchain's coinbase reward, a property that is emergent from the properties of the equilibrium and one that we exploit to increase security in Section~\ref{sec:security}. Altman et al.~\cite{Altman:2018} reached similar conclusions as Spiegelman et al.~\cite{Spiegelman:2018} using a slightly different game theoretical model of hash rate allocation across cryptocurrencies and mining pools.

Meshkov et al.~\cite{Meshkov:2017} introduced the term \emph{coin-hopping} to describe the strategy that involves some subset of miners moving among blockchains using the same PoW according to which is most profitable at a given time. They showed that this behavior can lead to unstable block times and proposed a modified difficulty adjustment algorithm to compensate. Coin-hopping corresponds to our definition of \emph{greedy behavior} in Section~\ref{sec:dynamics}. Király and Lomoschitz~\cite{Kiraly:2018} expanded on the study of the coin-hopping strategy, which they show can be profitable in the long-term for miners with at least 12\% of the total hash power. 

Han et al.~\cite{Han:2019} investigate doublespend on blockchains with relatively low hash rate instigated by either miners from a higher hash rate chain or attackers who purchase hash rate from a marketplace such as NiceHash\footnote{\url{https://www.nicehash.com}}. They find that doublespend transactions with fiat value on the order of 1e5 USD are sufficient to motivate Bitcoin miners to carry out an attack on Bitcoin Cash.

Several authors have sought to determine the optimal hash rate allocation between blockchains for \emph{individual} miners or mining pools. Bissias et al.~\cite{Bissias:2018} argue that miners allocate their hash rate between multiple blockchains so as to minimize the risk associated with fluctuations in coin price. Cong et al.~\cite{Cong:2018} make a similar argument except that their measure of risk is volatility in the payout rate between mining pools. Chatzigiannis et al.~\cite{Chatzigiannis:2019} extend this model to mining across blockchains with different PoW algorithms. All of the above approaches are complimentary to the present work,  which seeks only to explain \emph{aggregate} miner behavior. In fact, miner-specific behavioral choices help to explain why the aggregate hash rate allocation does not fully allocate to one chain over another (see Section~\ref{sec:greedy_cautious} for details).

Sapirshtein et al.~\cite{Sapirshtein:2015} devised a Markov Decision Process (MDP) for discovering optimal selfish mining~\cite{Eyal:2014} strategies. Gervais et al.~\cite{Gervais:2016} expanded the model to incorporate adjustable network parameters and include analysis of doublespend attacks. Sai et al.~\cite{Sai:2019} extend the MDP of Gervais et al.~\cite{Gervais:2016} to model mining difficulty adjustment. The biggest differences between these approaches and the present work is that the former analyze optimal \emph{deviant} behavior in single blockchains while the present work attempts to explain \emph{protocol compliant} behavior across multiple blockchains.

\section{Miner Allocation Among Blockchains}
\label{sec:dynamics}

In this section, we consider two blockchains $A$ and $B$, each generally using different PoW algorithms $W_A$ and $W_B$. Having different PoW algorithms, we imagine that the sets of miners $M_A$ and $M_B$ of each coin are generally disjoint, but in the special case where $W_A = W_B$ or when the algorithms support the same mining hardware, the sets can be equal or intersect. The \emph{native hash rate} (hashes per second) for miner $m$ is denoted $\mathcal{H}(m)$, and with $\mathcal{H}_A$ and $\mathcal{H}_B$ we denote the \emph{aggregate} native hash rate on chains $A$ and $B$, respectively. Through secondary markets such as NiceHash\footnote{\url{https://www.nicehash.com}}, an economically rational miner $m \in M_A$ will trade her hash power in $A$ for hash power in $B$ when the latter can earn her more fiat reward during the next moments of mining. Thus, miners $M_A \cup M_B$ collectively represent the aggregate achievable security of coins $A$ and $B$, which is fluid, subject to changes in the profitability of mining across chains $A$ and $B$.

\begin{mydef}
\label{def:spot_hashes}
The \emph{spot hash price} at time $s$, $\mathcal{S}_X(s)$, is the quantity of hashes per second using PoW algorithm $W_X$ that can be traded for 1 unit of fiat. 
\end{mydef}
	
Using definition~\ref{def:spot_hashes}, and assuming a perfectly efficient hash rate market, we can translate native hash rate on chain $B$ into units of native hash rate on chain $A$, a process we term \emph{hash rate regularization}. The regularized hash rates for chains $A$ and $B$ are equal to $H_A = \mathcal{H}_A$ and $H_B = \mathcal{H}_B \frac{\mathcal{S}_A}{\mathcal{S}_B}$. The regularized, aggregate hash rate across chains is given by $H = H_A + H_B$, where $H$ is native to chain $A$, a convention that we will follow throughout this document. By $H(m)$, we denote the regularized hash rate for miner $m$. Finally, define \emph{regularized allocation vector} $\boldsymbol{w} = (w_A, w_B)$ (or simply  \emph{allocation} for brevity) to be the fraction of $H$ that miners devote to chains $A$ and $B$, respectively. The following are definitions useful for discussing miner hash rate allocations and their relationship with blockchain security.

\begin{mydef}
The \emph{hash weight} of miner $m$, denoted $\mathcal{W}(m)$ is equal to $H(m) / H$, and the weight of a set of miners $M$, denoted $\mathcal{W}(M)$ is given by $\sum_{m \in M} \mathcal{W}(m)$.
\end{mydef}

\begin{mydef}
\label{def:rel_sec}
The \emph{relative security} of chain $X$ is the fraction of fiat value of PoW applied to that chain, which is given by
\begin{equation}
\label{eq:security}
K_X = \frac{\mathcal{H}_X / \mathcal{S}_X}{\mathcal{H}_A / \mathcal{S}_A + \mathcal{H}_B / \mathcal{S}_B} = \frac{H_X / \mathcal{S}_A}{H_A / \mathcal{S}_A + H_B / \mathcal{S}_A} = \frac{H_X}{H_A + H_B}.
\end{equation}
\end{mydef}

In terms of relative security, the regularized allocation vector is given by $\boldsymbol{w} = (K_A, K_B)$. Notice that the relative security for chain $X \in \{A, B\}$ is equivalent to the fraction of total available regularized hash rate (i.e., in terms of the $W_A$ PoW algorithm) allocated to chain $X$.  Thus, when chains $A$ and $B$ share the same PoW algorithm, $\boldsymbol{w}$ gives the share of hash rate for each chain.

The target, expected block inter-arrival time for chain $X \in \{A, B\}$ is denoted $T_X$. In general, blocks from chains $A$ and $B$ will be produced at different times, but we require some method of marking time universally. Let $\tau$ be a discrete variable that represents the times when a block is mined on either chain $A$ or $B$. At time $\tau$, the actual inter-arrival time for the last block from chain $X$ is given by $T_X(\tau)$, and the fiat coinbase value for chain $X$ is given by $V_X(\tau)$. Coinbase value decomposes into $V_X(\tau) = c_X P_X(\tau)$, where $c_X$ is the quantity of $X$ coins paid out per block and $P_X(\tau)$ is the fiat value of each $X$ coin at time $\tau$. Furthermore, define the \emph{hash adjusted reward}\footnote{The HAR is analogous to the popular \emph{difficulty adjusted reward index} (DARI) metric, except that the latter normalizes by the blockchain difficulty.} (\emph{HAR}) for chain $X$ at time $\tau$ by $\pi_X(\tau) = \frac{V_X(\tau)}{T_X(\tau) H_X(\tau)}$. The HAR for chain $X$ represents the expected fiat value of each regularized hash on chain $X$. Finally, define the relative reward of the two chains by $R(\tau) = \frac{V_A(\tau)}{V_A(\tau) + V_B(\tau)}$. Note that in this analysis we ignore the contribution of fees to the coinbase.

\begin{mydef}
A \emph{security adjustment algorithm} (SAA) is any algorithm that adjusts the expected number of  hashes required to mine blocks so that their expected inter-arrival time tends toward $T$; when the block time reaches $T$, the SAA is said to be \emph{at rest}. It is further assumed that the SAA is a function of the properties of previously mined blocks, thus it can only update the security after a new block is mined.\footnote{In practice, most PoW blockchains employ a \emph{difficulty adjustment algorithm}, which adjusts a value that is inversely proportional to the mining target $t$. Because \emph{difficulty} is ambiguously defined between blockchains, we opt for this definition instead.}
\end{mydef}

\begin{mydef}
The \emph{greedy choice} allocation, denoted $\boldsymbol{w}_g$, is one that yields that maximum weighted sum of HAR vector $\boldsymbol{\pi}(\tau) = (\pi_A(\tau), \pi_B(\tau))$, i.e. $\boldsymbol{w}_g^T \boldsymbol{\pi}$.
\end{mydef}

\begin{mydef}
Blockchains $A$ and $B$ are said to be at \emph{allocation equilibrium} if when both SAAs are at rest, there exists no greedy choice in allocation, i.e. $\boldsymbol{\pi} = c (1, 1)$ for some constant $c$.
\end{mydef}

\medskip

\begin{mylem}
\label{lem:dari}
Assume that at time $\tau$ both SAAs have come to rest, relative reward $R$ is stable, and the allocation vector is fixed at $\boldsymbol{w} = (xR, y(1 - R))$, where $x \geq 0$, $y \geq 0$, and $xR + y(1 - R) = 1$. Then the HAR vector is given by 
\begin{equation}
\label{eq:lem_dari}
\boldsymbol{\pi}(\tau) = \frac{V_A(\tau) + V_B(\tau)}{H} \left( \frac{1}{x T_A}, \frac{1}{y T_B} \right).
\end{equation}
\end{mylem}

\begin{myproof}
For $X \in \{A, B\}$, if at time $\tau$ the SAA for chain $X$ has come to rest, then the actual inter-block time $T_X(\tau)$ is approximately equal to its expected time $T_X$. Therefore, the HAR vector is given by
\[
\begin{array}{rcl}
(\pi_A(\tau), \pi_B(\tau)) 
& = & \left( \frac{V_A(\tau)}{T_A(\tau) H_A(\tau)}, \frac{V_B(\tau)}{T_B(\tau) H_B(\tau)} \right) \\
& = & \left( \frac{V_A(\tau)}{T_A H_A}, \frac{V_B(\tau)}{T_B H_B} \right) \\
& = & \frac{1}{H} \left( \frac{V_A(\tau)}{T_A w_A}, \frac{V_B(\tau)}{T_B w_B} \right) \\
& = & \frac{1}{H} \left( \frac{V_A(\tau)}{x T_A R}, \frac{V_B(\tau)}{y T_B (1-R)} \right) \\
& = & \frac{V_A(\tau) + V_B(\tau)}{H} \left( \frac{1}{x T_A}, \frac{1}{y T_B} \right).
\end{array}
\]
\end{myproof}

\bigskip
\begin{mythm}
\label{thm:equil}
Assume any choice of SAA for chains $A$ and $B$ (not necessarily the same), and further assume that total hash rate is fixed at $H$. When the relative reward stabilizes, there exists a unique equilibrium allocation 
\begin{equation}
\label{eq:equil}
\boldsymbol{w}_e = \left( \frac{T_B R}{T_B R - T_A R + T_A}, \frac{T_A (1 - R)}{T_B R - T_A R + T_A} \right),
\end{equation}
which simplifies to 
\begin{equation}
\boldsymbol{w}_e = (R, 1 - R),
\end{equation}
if $T_A = T_B$.
\end{mythm}

\bigskip
\begin{myproof}
An equilibrium allocation is one where the HAR vector is homogeneous, i.e. the HAR values for chains $A$ and $B$ are equal. Assuming the SAAs are at rest and relative price is stable, from Lemma~\ref{lem:dari} we can surmise that HAR values will be equal iff $x T_A = y T_B$. We can solve this equation for $x$ and $y$ along with the simultaneous constraint $xR + y(1 - R) = 1$:
\[
x_e = \frac{T_B}{T_B R - T_A R + T_A}, \text{~and~} y_e = \frac{T_A}{T_B  R - T_A R + T_A}.
\]
Substituting $x_e$ and $y_e$ into the identity $\boldsymbol{w}_e = (x_e R, y_e (1 - R))$ (from the statement of Lemma~\ref{lem:dari}) yields Equation~\ref{eq:equil}, as desired. Moreover, because the constraints constitute a system of two linearly independent equations with two unknowns, $\boldsymbol{w}_e$ must be the only equilibrium allocation.
\end{myproof}

We next derive results related to how miners behave relative to the equilibrium.

\begin{mydef}
\label{def:distance}
The \emph{distance} between two allocations $\boldsymbol{w}_1$ and $\boldsymbol{w}_2$ is given by the L1-norm of their difference: $|\boldsymbol{w}_1 - \boldsymbol{w}_2|$.
\end{mydef}

\begin{mydef}
\label{def:eps_greedy}
The \emph{$\epsilon$-greedy allocation policy} moves the current allocation $\epsilon$ closer (in terms of Definition~\ref{def:distance}) to the greedy choice, e.g. if $\pi_A(\tau_0) > \pi_B(\tau_0)$, then $\boldsymbol{w}(\tau_1) = (w_A(\tau_0)+\frac{\epsilon}{2}, w_B(\tau_0)-\frac{\epsilon}{2})$.
\end{mydef}

\begin{mydef}
The set of miners \emph{loyal}\footnote{Our definition of \emph{loyal} is consistent with Király and Lomoschitz~\cite{Kiraly:2018}, but not Kwon et al.~\cite{Kwon:2019}.} to chain $X \in \{A, B\}$, denoted by $M_{X^*}$, are those that will allocate all hash rate to chain $X$ over chain $Y$ regardless of the value of $\pi_X$ relative to $\pi_Y$.
\end{mydef}

\bigskip

\begin{mythm}
\label{thm:equil_convg}
Assume that equilibrium allocation $\boldsymbol{w}_e$ is fixed over an arbitrarily long period of time and loyal miner hash weights are such that $\mathcal{W}(M_{A^*}) \leq w_{eA} H$ and $\mathcal{W}(M_{B^*}) \leq w_{eB} H$. If the SAAs on both chains have come to rest and reward ratio $V_A(\tau) / V_B(\tau)$ is constant in $\tau$, then from an allocation not at equilibrium and for sufficiently small $\epsilon$, the $\epsilon$-greedy allocation policy converges to the equilibrium allocation.
\end{mythm}

\begin{myproof}
We prove this result in two stages. In the first we show that a non-loyal miner, making an $\epsilon$-greedy choice will always move in the direction of $\boldsymbol{w}_e$. In the second, we argue that non-loyal miners comprise sufficient hash weight to reach $\boldsymbol{w}_e$.

To prove the first stage, it will suffice to show that for a suitably small $\epsilon$, the $\epsilon$-greedy allocation $\boldsymbol{w}(\tau_1)$ always moves the current allocation $\boldsymbol{w}(\tau_0)$ closer to the equilibrium allocation $\boldsymbol{w}_e$. Without loss of generality, we may assume that $\pi_A(\tau_0) > \pi_B(\tau_0)$. In this case, because $w_B(\tau_0) = 1 - w_A(\tau_0)$, we need only show that $w_A(\tau_0) < w_{eA}$ to prove the theorem. This follows from the fact that our assumption implies that the greedy choice will increase $w_A$: $w_A(\tau_1) = w_A(\tau_0) + \frac{\epsilon}{2}$, which can only move the allocation closer to $w_{eA}$ provided that $\frac{\epsilon}{2} < w_{eA} - w_A(\tau_0)$.

Before proceeding, note that because the reward ratio is stable, there exists an $r$ such that $V_A(\tau) / V_B(\tau) = r$ for all $\tau$. Similarly, because the SAAs are assumed to have come to rest, we assume that $T_X(\tau) = T_X$ for every $\tau$ and $X \in \{A, B\}$. We have
\[
\begin{array}{rcl}
\pi_A(\tau_0) > \pi_B(\tau_0) 
& \Rightarrow & \frac{V_A(\tau_0)}{H_A(\tau_0) T_A} > \frac{V_B(\tau_0)}{H_B(\tau_0) T_B} \\
& \Rightarrow & \frac{V_A(\tau_0)}{H w_A(\tau_0) T_A} > \frac{V_B(\tau_0)}{H w_B(\tau_0) T_B} \\
& \Rightarrow & \frac{a}{w_A(\tau_0)} > \frac{b}{1 - w_A(\tau_0)} \\
& \Rightarrow & w_A (\tau_0) < \frac{a}{a+b},
\end{array}
\]
where $a = r / T_A$ and $b = 1 / T_B$. On the other hand, similar reasoning shows that $w_{eA} = \frac{a}{a+b}$. So we have $w_A(\tau_0) < w_{eA}$, as required.

To prove the second stage, it will suffice to argue that the hash weight of non-loyal miners at time $\tau_0$ is non-zero. We again assume without loss of generality that that $\pi_A(\tau_0) > \pi_B(\tau_0)$. Let $M = M_A \cup M_B$, and note that by definition $\mathcal{W}(M) = H$. The hash weight of non-loyal miners is given by $\mathcal{W}M) - \mathcal{W}(M_{A^*}) - \mathcal{W}(M_{B^*})$. In stage~1, we proved that $w_A(\tau_0) < w_{eA}$, which implies that $\mathcal{W}(M_{A^*}) < w_{eA} H$. And by assumption $\mathcal{W}(M_{B^*}) \leq w_{eB} H$. Finally, because $w_A(\tau_0) + w_B(\tau_0) = 1$, we know that $\mathcal{W}(M) - \mathcal{W}(M_{A^*}) - \mathcal{W}(M_{B^*}) > 0$, which implies that the hash weight of non-loyal miners at time $\tau_0$ must be non-zero.
\end{myproof}

\begin{mycor}
\label{cor:divergence}
If the SAAs on both chains have come to rest and reward ratio $V_A / V_B$ is stable, then for any allocation within distance $\delta$ of the equilibrium allocation, following the $(2 \delta + \epsilon)$-greedy allocation policy, $\epsilon > 0$, causes divergence from the equilibrium allocation. 
\end{mycor}

\begin{myproof}
Similar to the proof of stage~1 in Theorem~\ref{thm:equil_convg}, it will suffice to show that for any $\epsilon > 0$, the $(2 \delta+\epsilon)$-greedy allocation $\boldsymbol{w}(\tau_1)$ always moves the current allocation $\boldsymbol{w}(\tau_0)$ \emph{further} from the equilibrium allocation $\boldsymbol{w}_e$. Let $\delta_A = |w_A(\tau_0) - w_{eA}|$ and $\delta_B = |w_B(\tau_0) - w_{eB}|$, which according to Definition~\ref{def:distance} must satisfy $\delta_A + \delta_B = \delta$. Again, without loss of generality, we may assume that $\pi_A(\tau_0) > \pi_B(\tau_0)$, which implies that the greedy choice will increase $w_A$: $w_A(\tau_1) = w_A(\tau_0) + \delta + \frac{\epsilon}{2}$. 
The proof of stage~1 of Theorem~\ref{thm:equil_convg} showed that $w_A(\tau_0) < w_{eA}$, giving $w_{eA} = w_A(\tau_0) + \delta_A$. It follows then that 
\[
w_A(\tau_1) = w_A(\tau_0) + \delta + \frac{\epsilon}{2} = w_{eA} - \delta_A + \delta + \frac{\epsilon}{2}.
\]
Similarly, $w_B(\tau_1) = w_{eB} + \delta_B - \delta - \frac{\epsilon}{2}$. Therefore, the $(2\delta + \epsilon)$-greedy choice at time $\tau_0$ moves $\boldsymbol{w}(\tau_1)$ further from $\boldsymbol{w}_e$ by $|w_A(\tau_1) - w_{eA}| + |w_B(\tau_1) - w_{eB}| = |\delta -\delta_A + \epsilon / 2| + |\delta_B -\delta + \epsilon / 2| = \epsilon$.
\end{myproof}

\begin{mydef}
The \emph{extreme greedy} policy for a non-loyal miner is to allocate all hash rate entirely to the greedy choice.
\end{mydef}

\begin{mycor}
\label{cor:oscillations}
Let $\mathcal{W}(M_{A^*})$ and $\mathcal{W}(M_{B^*})$ be the hash weights of miners loyal to coins $A$ and $B$, respectively, and define
\[
\boldsymbol{w}_1 = (1-\mathcal{W}(M_{B^*}), \mathcal{W}(M_{B^*})) \frac{1}{H} \text{~and~} \boldsymbol{w}_2 = (\mathcal{W}(M_{A^*}), 1-\mathcal{W}(M_{A^*})) \frac{1}{H}.
\]
Suppose equilibrium allocation $\boldsymbol{w}_e$ is such that $w_{2A} < w_{eA} < w_{1A}$ and $w_{1B} < w_{eB} < w_{2B}$, and suppose further that reward ratio $V_A(\tau) / V_B(\tau)$ is constant in $\tau$. Then for any choice of SAAs, non-loyal miners following the extreme greedy policy will result in hash rate fluctuations that oscillate between $\boldsymbol{w}_1$ and $\boldsymbol{w}_2$. SAAs that come to rest faster will result in higher frequency oscillations.
\end{mycor}

\begin{myproof}
Without loss of generality, we can assume that at time $\tau_0$ the greedy choice is to allocate all hash rate to chain $A$, which implies that $\boldsymbol{w}(\tau_0) = \boldsymbol{w}_1$. Now assume that both SAAs have come to rest at time $\tau_1$ (if $\mathcal{W}(M_{B^*}) = 0$, then the SAA for chain $B$ will not have had an opportunity to run because it has hash rate zero, but we nominally regard this as being at rest). It will suffice to show that the greedy choice at time $\tau_1$ is to shift allocation to coin $B$. Suppose, for the purpose of contradiction, that the greedy choice at time $\tau_1$ is to maintain maximum allocation to coin $A$. In that case, according to Theorem~\ref{thm:equil_convg}, there must exist some $\epsilon > 0$ such that $|w_{1A} + \frac{\epsilon}{2} - w_{eA}| < |w_{1A} - w_{eA}|$. But this is not possible because, by assumption $w_{1A} > w_{eA}$, so the greedy choice at time $\tau_1$ must instead be to shift allocation to coin $B$,  i.e. $\boldsymbol{w}(\tau_1) = \boldsymbol{w}_2$. Notice that SAAs that come to rest faster will realize faster fluctuations in the extreme greedy choice, and will therefore result in higher frequency oscillations between extreme allocations.
\end{myproof}

\section{Beyond PoW}
\label{sec:beyond_pow}

Fundamentally, the results of Section~\ref{sec:dynamics} tie the aggregate relative security of a blockchain to the value of reward given to those who provide security (i.e., PoW). PoW can be seen as proof-of-opportunity-cost for miners, who sacrifice energy and CPU cycles in return for the opportunity to gain native coins and a \emph{vote} on the next block. The HAR measures fiat value per unit of opportunity cost. And the SAA is simply a means of regulating this value so as to achieve to the desired emission of the native currency. 

We can generalize PoW concepts as follows. Each blockchain defines a \emph{cost function} $\mathcal{C}_X$ with which it maps a unit of \emph{native cost} to some quantity of native coin $X$. In PoW, native cost for chain $X$ is the execution of a single hash using algorithm $W_X$. Define a \emph{proof of cost} (PoC) voting system as one that allocates votes and native reward to participants proportional to their demonstrated cost. Furthermore, define a \emph{cost adjustment algorithm} (CAA) as an algorithm that adjusts cost function $\mathcal{C}_X$ so as to achieve a desired distribution of coin $X$ over the short-to-medium-term. Total cost per second, $H_X$, on chain $X$ is the amount of cost levied collectively against all participants in a single second. Regularized hash rate, $\mathcal{H}_X$ is interpreted as the total cost per second on chain $X$, denominated in units chain $A$ cost. The cost-adjusted-reward (CAR) is the fiat value of reward per unit of regularized cost. Some proof-of-stake (PoS) systems meet the criteria of a PoC voting system, and therefore, there exists the potential for an equilibrium to form relative to a PoW blockchain.

Public blockchains produce blocks as the result of a voting process, where votes are awarded to participants proportional to their opportunity cost. In PoW systems, the set of participants is entirely \emph{open}: anyone with access to hardware capable of running the PoW algorithm can vote. But in PoS system, the set of participants is \emph{restricted}: only those holding native coins can vote. Moreover, most PoS systems make a distinction between \emph{active validators} who actively stake coins and simple coin \emph{owners}. The former set can vote, while the latter cannot. Blocks are produced in \emph{validation rounds}. Delegated PoS or DPoS blockchains are somewhat different still; coin holders vote for \emph{delegates} and it is the delegates that create blocks using an alternative form of consensus such as Byzantine Fault Tolerance~\cite{Lamport:1982}.  Another difference is that, instead of fixing the number of coins $c$ comprising the block reward, PoS blockchains tend to define $c$ as a function of the number of coins staked by active validators. 

\subsection{Basic PoS Equilibria}

Consider PoS blockchain $X$. At time $\tau$, there are $k_X$ coins staked on chain $X$. The total reward for a single validation round is $c_X$, and each round lasts $T_X$ seconds. Thus, total reward value is given by $V_X = c_X P_X$, where $P_X$ is the fiat value of coin $X$. The total opportunity cost during a validation round, $\mathcal{H}_X$, is equal to $r k_X T_X P_X$, where $r$ is the risk-free rate of return for investing 1 unit of fiat for 1 second. In words, $\mathcal{H}_X$ measures the amount of fiat that could be earned by exchanging quantity $k_X$ coins $X$ for a so-called risk-free asset such as the 1-year US Treasury Note. Because the \emph{native} unit of cost for $\mathcal{H}_X$ is fiat, $\mathcal{S}_X = 1$, and $\mathcal{H}_X \mathcal{S}_A = H_X$. The CAR is given by $\pi_X = \frac{V_X}{H_X}$. 

\bigskip

\begin{myexam}
\label{exam:neo}
NEO is a DPoS blockchain~\cite{NEO}. There are two native coins on the chain: NEO and GAS. Holding NEO affords the bearer two privileges: the right to vote for delegates and access to a stream of GAS. Exactly 100e6 NEO coins exist; initially 50e6 were distributed during a crowd sale and the remaining 50e6 were reserved by the \emph{NEO council} to be used in the future to pay for development. GAS is awarded to NEO holders every validation round (occurring roughly once every $T_{\text{GAS}}$ seconds) according to their percentage of the total available NEO.
Initially, 8 GAS per round where awarded total, i.e. $c_{\text{GAS}} = 8$. Every 2e6 validation rounds (roughly 1 year), $c_{\text{GAS}}$ is reduced by 1 GAS\footnote{\url{https://docs.neo.org/docs/en-us/basic/whitepaper.html}}. 
As of July 27, 2019, more than 4e6 blocks have been mined\footnote{\url{https://neotracker.io}}, therefore $c_{\text{GAS}} = 6$. With these definitions, we can directly compare the security of the NEO blockchain to that of an arbitrary PoW blockchain using the framework from Section~\ref{sec:dynamics}.
\end{myexam}

\section{Evaluation}
\label{sec:eval}

\begin{figure}[t]
\includegraphics[width=\linewidth]{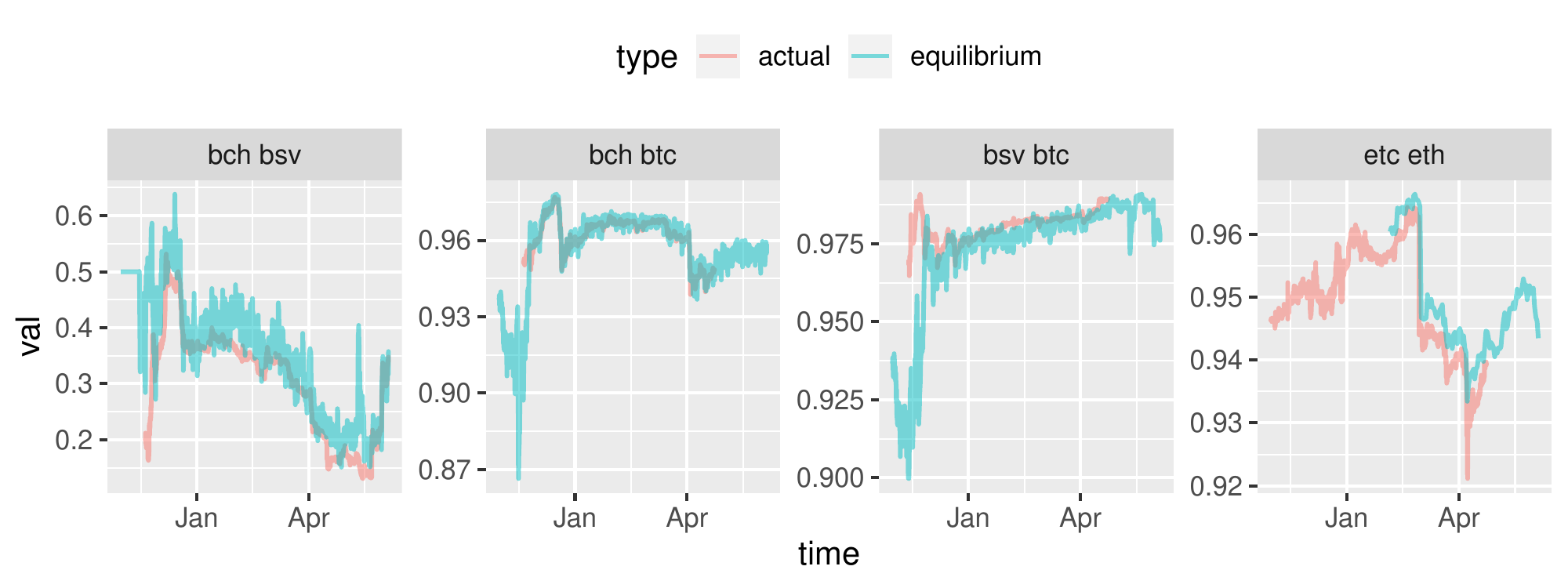}
\caption{Actual hash rate allocation between various pairs of cryptocurrencies (red) juxtaposed with the equilibrium allocation (blue). The plots show strong agreement between the actual allocation and the allocation predicted by the equilibrium, the latter of which is based entirely on expected block times and coinbase values. The data ranges from December 1, 2018 until June 1, 2019.}
\label{fig:actual_vs_equilibirum}
\end{figure}

In this section, we validate the theoretical results from Section~\ref{sec:dynamics}. Recall that the \emph{actual} resource allocation between two blockchains is given by $\boldsymbol{w} = (K_A, K_B)$, where $K_X$ is the relative security of chain $X \in \{A, B\}$ (see Definition~\ref{def:rel_sec}). When chains $A$ and $B$ employ the same PoW algorithm $W$, $\boldsymbol{w}$ is simply the fraction of aggregate hash rate for algorithm $W$ applied to each of the chains. (Note that by \emph{aggregate} we mean between chains $A$ and $B$ only, so $|\boldsymbol{w}| = 1$.) We first show that the equilibrium point $\boldsymbol{w}_e$, described by Theorem~\ref{thm:equil}, closely matches the actual allocation $\boldsymbol{w}$ for real historical blockchain data. We then show results from a block mining simulation that corroborate Theorem~\ref{thm:equil_convg} and Corrolary~\ref{cor:oscillations}.

\subsection{Historical Convergence to Equilibrium}

We collected historical data for several of the largest PoW blockchains by market cap including Bitcoin (BTC), Bitcoin Cash (BCH), Bitcoin Satoshi Vision (BSV), Ethereum (ETH), and Ethereum Classic (ETC). Included in the dataset were hourly prices for ETH, ETC, and BSV from the Bitfinex exchange\footnote{\url{https://www.bitfinex.com}} and for BTC and BCH from the Binance exchange\footnote{\url{https://www.binance.com}}. Difficulty data were collected for each block from Blockchair\footnote{\url{https://blockchair.com}} with the exception of ETC, which was collected hourly from Coinwarz\footnote{\url{https://coinwarz.com}}. From the difficulty, we were able to extract the approximate hash rate for each of the blockchains. Figure~\ref{fig:actual_vs_equilibirum} plots the actual hash rate allocation $\boldsymbol{w}$ in red for various pairs of blockchains (one pair per facet) along with the equilibrium allocation $\boldsymbol{w}_e$, which is plotted in blue. Agreement between the two curves indicates that there was an observed convergence to equilibrium as predicted by Theorem~\ref{thm:equil_convg}. The plots generally show strong agreement except for times when there were well-documented macro-level disturbances. For example, BSV hard-forked from BCH in November, 2018\footnote{\url{https://en.wikipedia.org/wiki/Bitcoin\_Cash\#November\_2018\_split}}, and we can see corresponding divergences from the equilibrium at this time. Also, a bug was exploited in the BCH ABC client during a hard fork upgrade in May, 2019, which caused a delay in block production and a chain reorganization\footnote{\url{https://cointelegraph.com/news/bitcoin-cash-experiences-bug-during-scheduled-hard-fork-upgrade}}. Corresponding to this event, we again see divergence from the equilibrium. 

There are two reasons why allocations might diverge from equilibrium at these times. First, the equilibrium defined by Theorem~\ref{thm:equil} assumes that blocks arrive exactly at their targeted times (every 10 minutes for BCH and BSV). However, during the events discussed above, block times were significantly slower than 10 minutes for a period of time, which means the plotted equilibrium is not quite accurate. Second, prices tend to fluctuate wildly during hard-forks and when bugs are encountered. (Indeed, there was no trading of BCH or BSV for several days on most exchanges during the November, 2018 hard-fork.) At those times it is difficult to correctly formulate an equilibrium with inaccurate price data, and also it is possible that miners will cease to mine greedily in order to ensure a given chain continues to produce blocks.

\subsection{Convergence to Equilibrium in Simulation}

\begin{figure}[t]
\begin{minipage}[c]{0.63\textwidth}
\includegraphics[width=\linewidth]{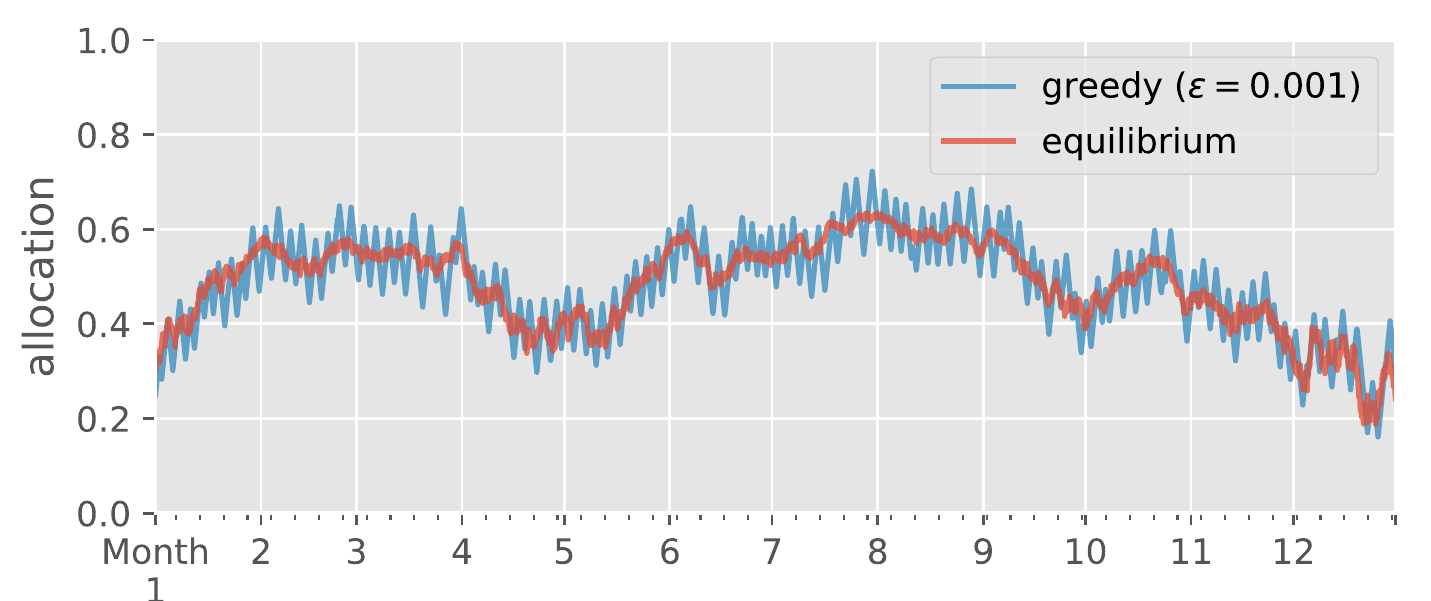}
\includegraphics[width=\linewidth]{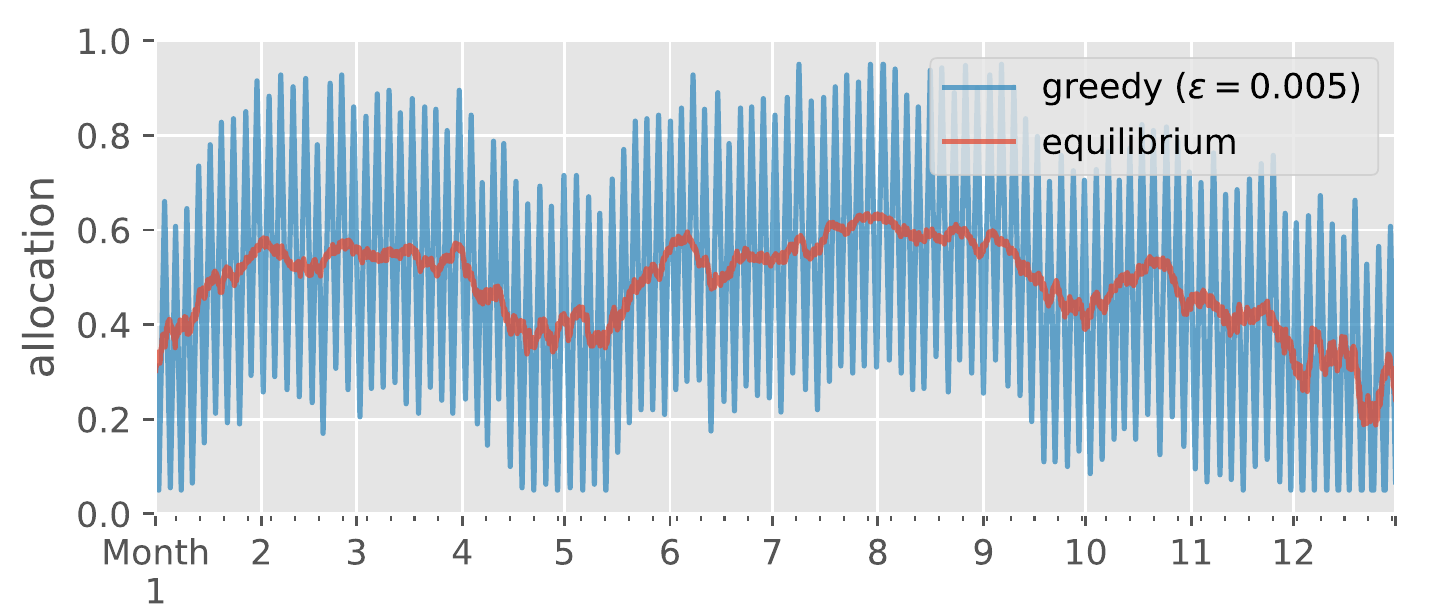}
\includegraphics[width=\linewidth]{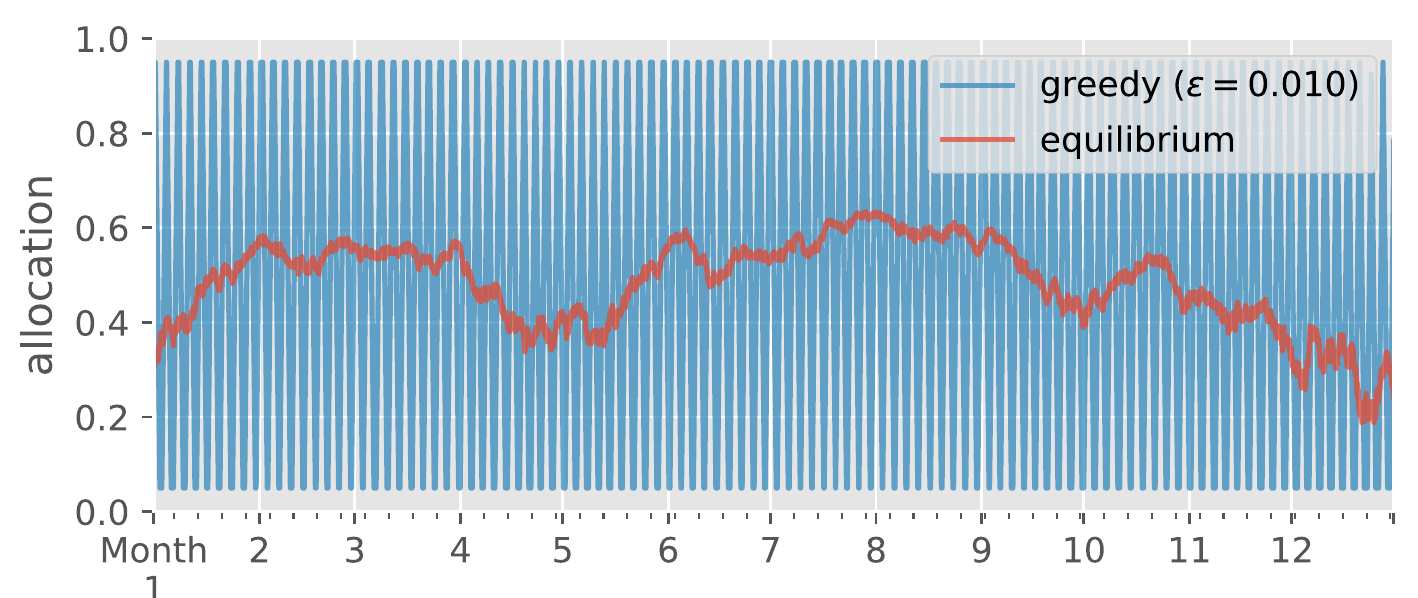}
\end{minipage}
\begin{minipage}[c]{0.35\textwidth}
\caption{Results from 3 block mining simulation runs where miners chose between chains $A$ and $B$. Each curve shows the equilibrium (red) and actual (blue) allocations to chain $B$. Each facet varies $\epsilon$ and 90\% of miners follow the $\epsilon$-greedy policy. Of the remaining miners, 5\% are loyal to chain $A$ and 5\% are loyal to chain $B$. The plots show close adherence to the equilibrium for small $\epsilon$ and wild oscillations between extreme allocations as $\epsilon$ increases.}
\label{fig:actual_vs_equilibirum}
\end{minipage}\vspace{-2ex}
\end{figure}

We implemented a block mining simulation where miners were given the choice between chain $A$ or $B$. For simplicity, it was assumed that both chains used the same PoW algorithm, had the same target inter-block time, and issued the same number of coins per block. The ratio of prices was initially 0.5, but we allowed it to vary according to a random walk with mean 0 and standard deviation 5e-3. We simulated 15 months of block generation total, but discarded the first 90 days to ensure the system had reached a steady state. Both chains $A$ and $B$ were assumed to use the difficulty adjustment algorithm (DAA) of Bitcoin Cash (BCH)~\cite{Sechet:2017}, which uses a rolling average of the ratio of chain work to block time over the last 144 blocks (roughly 1 day). For each run of the simulation, we assumed that 5\% of the miners were loyal to chain $A$, 5\% were loyal to chain $B$, and the remaining 90\% followed the $\epsilon$-greedy strategy where $\epsilon$ was allowed to vary between runs. 

Figure~\ref{fig:actual_vs_equilibirum} shows the results of a single simulation run for each choice of $\epsilon \in \{\text{1e-3, 5e-3, 1e-2}\}$. The plots show the second component of the equilibrium vector ($w_{eB}$) in red, juxtaposed with the actual aggregate allocation to chain $B$ in blue. As predicted by Theorem~\ref{thm:equil_convg}, sufficiently small $\epsilon$ (top facet) ensures convergence to the equilibrium. On the other hand, larger choices for $\epsilon$ result in divergence from the equilibrium (lower two facets), as predicted by Corollary~\ref{cor:divergence}. Moreover, as predicted by Corollary~\ref{cor:oscillations}, the divergence from the equilibrium results in oscillations between the extremes defined by the fraction of loyal miners: $\mathcal{W}(M_{B^*})$ at one extreme and $1 - \mathcal{W}(M_{A^*})$ at the other.

\section{Applications}

\subsection{Trustless Price-Ratio Oracle}
\label{sec:oracle}

Price feeds are a fundamental tool for many popular smart contract applications including prediction market Augur~\cite{Augur}, stable coin issuer MakerDAO~\cite{Makerdao}, hedge fund Numerai~\cite{Numerai}, and loan initiator Dharma~\cite{Dharma}. Existing price feed solutions range from crowd-sourced~\cite{whgeorge,DutchX} to trusted/whitelisted sources~\cite{MakerdaoPrice,Provable}. In this section, we present an application of the allocation equilibrium presented in Section~\ref{sec:dynamics} that delivers an estimate of the fiat \emph{price ratio} of the coins native to blockchains $A$ and $B$ who share the same PoW algorithm.

We now describe how chain $A$ can implement a price ratio oracle, but it should be noted that chain $B$ could also do the same. Smart contract $\texttt{Oracle}$ runs on chain $A$ and returns an estimate of the price ratio $P_B / P_A$ when the two chains are each at a given block height. Essentially, $\texttt{Oracle}$ runs a light client for blockchain $B$, which contains all the headers since the chain's genesis block. There are just two public methods exposed: $\texttt{Update}(h_B)$ and  $\texttt{Query}(b_A, b_B)$. Method $\texttt{Update}(h_B)$ allows any user or other contract to update the chain of headers with a new header $h_B$ having the following properties: \1 the block height of $h_B$ is exactly one greater than the previous header; \2 the previous block hash of $h_B$ points to the block hash of the previous header; and \3 the PoW associated with the the hash of $h_B$ meets the difficulty implied by earlier headers and chain $B$'s protocol. If any of the conditions are not met, then it returns an error. Method $\texttt{Query}(b_A, b_B)$ returns an estimate of the price ratio $P_B / P_A$ at the time when chain $A$ was at block height $b_A$ and chain $B$ was at height $b_B$. If either \1 the header at block height $b_B$ is unknown to $\texttt{Oracle}$ or \2 the block on chain $A$ at height $b_A$ has not yet been mined, then an error is returned. Any party interested in maintaining the validity of the oracle will be sure to quickly run $\texttt{Update}(h_B)$ for all new headers $h_B$ for chain $B$. 

We assume that $\texttt{Oracle}$ will have native access to the header of the current block on chain $A$, $h_A$. Using headers $h_A$ and $h_B$, $\texttt{Oracle}$ will estimate $P_B / P_A$ using Definition~\ref{def:spot_hashes} and Theorem~\ref{thm:equil}. Let $H(h_X)$ be an estimate of the hash rate for chain $X$ derived from header $h_X$. On Bitcoin-like blockchains, the hash rate is simply $\sfrac{2^{32}}{D}$ where $D$ is the difficulty, which is included in the block header. Hash rate can be extracted via similar transformations on other blockchains. From Definition~\ref{def:rel_sec}, we have that 
\[
w_A \approx \frac{H(h_A)}{H(h_A) + H(h_B)},
\]
where $w_A$ denotes the fraction of total hash rate $H$ shared between chains $A$ and $B$ that is devoted to chain $A$. According to Theorem~\ref{thm:equil}, 
\[
w_A = \frac{T_B R}{T_B R - T_A R + T_A}
\] 
at equilibrium, where $T_A$ and $T_B$ are expected block times for chains $A$ and $B$, $R = \frac{V_A}{V_A + V_B}$, $V_X = c_X P_x$, and $c_X$ is the number of coins rewarded per block on chain $X \in \{A, B\}$. Therefore,
\[
\renewcommand*{\arraystretch}{2.0}
\begin{array}{l}
\frac{H(h_A)}{H(h_A) + H(h_B)} \approx \frac{T_B R}{T_B R - T_A R + T_A} \Rightarrow \\
\frac{P_B}{P_A} \approx \frac{c_A}{c_B T_A} \left( \frac{T_B (H(h_A) + H(h_B)) }{H(h_A)} - T_B + T_A \right) - \frac{c_A}{c_B}.
\end{array}
\]

Figure~\ref{fig:actual_vs_equilibirum} demonstrates that the equilibrium (shown in blue) typically agrees strongly with the security implied by the relative hash rate (shown in red). For this reason, we expect that price-ratio predictions will often be quite good. Of course an approximation of this sort is likely never to be as good as a centralized price feed. Thus, we envision the consumers of $\texttt{Oracle}$ to be users or smart contracts that require a fully decentralized oracle, or perhaps require a safety check on the trust placed in a centralized price feed.

\bigskip

\begin{myexam}
\label{exam:futures}
Suppose that we wish to introduce fully decentralized \emph{futures contracts} to blockchain $A$ intended to be negotiated between two parties: guarantor $\mathcal{G}$ and beneficiary $\mathcal{B}$. To do so, a smart contract can be developed that leverages $\texttt{Oracle}$. Each futures contract, or \emph{future} transfers from guarantor to beneficiary a quantity of coins $A$ equivalent to the value of a quantity of coin $B$ at a future date. 
Specifically, a future issued at the time when chains $A$ and $B$ are at block heights $b_A$ and $b_B$, allows the beneficiary to trade the contract to the guarantor for a quantity of coins $A$ equivalent to 1 coin $B$ on the \emph{expiry date}. We define expiry as the latter of block heights $b'_A$ and $b'_B$, anticipated to be some time in the future (for example 90 days). Contract $\texttt{Future}$ implements four methods: $\texttt{Deposit}(a)$, $\texttt{Recover}(a)$, $\texttt{Issue}(b_A, b_B, b'_A, b'_B, a)$, and $\texttt{Redeem}(b_A', b_B')$. $\texttt{Deposit}$ is signed by $\mathcal{G}$; it deposits quantity $a$ of coin $A$ into $\texttt{Future}$. This will be used to pay $\mathcal{B}$ at expiry. Prior to calling $\texttt{Issue}$, the funds can be redeemed by $\mathcal{G}$ if he signs $\texttt{Recover}$. The call to $\texttt{Issue}$ must be signed by both $\mathcal{G}$ and $\mathcal{B}$; signifying that they agree to the initial and expiry block times and fee of $a$ coins, which is paid by $\mathcal{B}$ and immediately transferred to an account owned by $\mathcal{G}$. Once headers $h'_A$ and $h'_B$ at height $b'_A$ and $b'_B$ have been generated, $\mathcal{B}$ first calls $\texttt{Update}(h'_B)$ on $\texttt{Oracle}$ and then signs $\texttt{Redeem}$. In response to this method, contract $\texttt{Future}$ deposits into an account controlled by $\mathcal{B}$ a quantity of $A$ coins that are equivalent to the value of 1 coin $B$ as determined by calling $\texttt{Query}(b_A', b_B')$ on contract $\texttt{Oracle}$. 
\end{myexam}

\subsection{Increasing Security}
\label{sec:security}

Consider two blockchains $A$ and $B$ that have the same PoW algorithm $W$ and target inter-block time $T$. BTC and BCH constitute an example where $W$ is the \texttt{SHA256} algorithm and $T = 600$ seconds. Recall from Section~\ref{sec:dynamics} that $P_A$ and $P_B$ are the fiat coin values for $A$ and $B$, respectively, and that, ignoring fees, coinbase value $V_X = c_X P_X$, where $c_X$ denotes the number of coins issued per block on chain $X$. Finally, recall that $R = V_A / (V_A + V_B)$. Suppose that coin $B$ is consistently less valuable than coin $A$; i.e., $P_B / P_A = \alpha$ for some $\alpha < 1$. If $c_A = c_B$, then because they share the same PoW algorithm and inter-block time, Theorem~\ref{thm:equil} predicts that the equilibrium allocation will be 
\begin{equation}
\label{eq:equil_alpha}
\begin{array}{rcl}
\boldsymbol{w}_e &=& (R, 1-R) \\ 
&=& \left(\frac{V_A}{V_A+V_B}, \frac{V_B}{V_A+V_B} \right) \\
&=& \left(\frac{P_A}{P_A+P_B}, \frac{P_B}{P_A+P_B} \right) \\
&=& \frac{1}{1 + \alpha} (1, \alpha).
\end{array}
\end{equation}
Thus, chain $A$ will tend toward having $1 / \alpha$ more hash rate than chain $B$, which constitutes lower security for chain $B$. This can lead to a negative feedback loop where lower security leads to lower coin price, which in turn leads to even lower security. One way to break this loop is for chain $B$ to simply increase the issuance per block, $c_B$. Of course $P_B$ will be reduced in value as a result, but somewhat surprisingly, the net effect is not necessarily zero sum.

The \emph{market capitalization} (CAP) for coin $X$, $m_X$, is a measure of the aggregate future value of the corresponding blockchain in the same sense that the CAP of an equity is a measure of the capacity for the underlying corporation to deliver returns to investors in the future. We do not attempt to economically justify the CAP of blockchain coins, but rather we treat the CAP as an objective measure of overall blockchain value that is emergent from the coin market. At time $\tau$, CAP is related to circulating coin $I_X$ and coin price by $m_X = I_X(\tau) P_x(\tau)$. Because no new value is generated for a blockchain by circulating more coin, an increase in issuance alone should not increase the CAP. But since more coin has been issued, the fiat price of each coin must decrease. Therefore, an increase of $\Delta I$ coins for chain $B$ during time $\Delta \tau$ must decrease the value of coin $B$ by
\begin{equation}
\label{eq:issue_price}
\begin{array}{rcl}
\Delta P_B &=& m_B \left( \frac{1}{I_B(\tau) + \Delta I} - \frac{1}{I_B(\tau)} \right) \\
&=& \frac{m_B}{I_B(\tau) + \Delta I} \left( 1 - \frac{I_B(\tau) + \Delta I}{I_B(\tau)} \right) \\
&=& - \frac{m_B}{I_B(\tau) + \Delta I} \frac{\Delta I}{I_B(\tau)} \\
&=& - \frac{P_B(\tau) \Delta I}{I_B(\tau) + \Delta I} \\
\end{array}
\end{equation}

Suppose that until time $\tau$, chains $A$ and $B$ have each issued $I_A(\tau) = I_B(\tau) = I$ total coins and have each issued the same number of new coins per block: $c_A(\tau)= c_B(\tau) = c$. Because $P_B / P_A = \alpha$, we also have $m_B / m_A = \alpha$.
At time $\tau$, chain $B$ decides to increase its issuance per block by factor $k > 1$ for a period of time $\Delta \tau$, i.e. $c_B(\tau') = kc$ while $c_A(\tau') = c$ for $\tau' \in [\tau, \tau + \Delta \tau]$. At time $\tau + \Delta \tau$, $\beta I$ total $A$ coins and $\gamma I$ total $B$ coins will have been issued, where $\gamma > \beta > 1$. As a result, according to Equation~\ref{eq:issue_price}, 
\[
\begin{array}{rcl}
P_A(\tau + \Delta \tau) &=& P_A(\tau) - \frac{(\beta-1) I P_A(\tau)}{\beta I} \\
&=& P_A(\tau) \left(1 - \frac{\beta - 1}{\beta} \right) \\
&=& \frac{1}{\beta} P_A(\tau).
\end{array}
\]
while
\[
\begin{array}{rcl}
P_B(\tau + \Delta \tau) &=& P_B(\tau) - \frac{(\gamma-1) I P_B(\tau)}{\gamma I} \\
&=& P_B(\tau) \left(1 - \frac{\gamma - 1}{\gamma} \right) \\
&=& \alpha P_A(\tau) \left(1 - \frac{\gamma - 1}{\gamma} \right) \\
&=& \frac{\alpha}{\gamma} P_A(\tau).
\end{array}
\]
At time $\tau + \Delta \tau$, noting that $R = P_A / (P_A + k P_B)$, the equilibrium allocation becomes
\begin{equation}
\label{eq:equil_incr}
\begin{array}{rcl}
\boldsymbol{w}_e(\tau + \Delta \tau) &=& (R, 1-R) \\ 
&=& \left(\frac{P_A}{P_A+kP_B}, 1 - \frac{P_A}{P_A+k P_B} \right) \\
&=& \left( \frac{\frac{1}{\beta} }{\frac{1}{\beta} + \frac{k \alpha}{\gamma}}, 1 - \frac{\frac{1}{\beta} }{\frac{1}{\beta} + \frac{k \alpha}{\gamma}} \right) \\
&=& \left(\frac{1}{1 + \frac{k \alpha \beta}{\gamma}}, \frac{\alpha}{\alpha + \frac{\gamma}{k \beta}} \right).
\end{array}
\end{equation}
Comparing the new equilibrium in Equation~\ref{eq:equil_incr} to the equilibrium prior to the increase in issuance given by Equation~\ref{eq:equil_alpha}, we see that chain $B$ will have increased its share of the hash rate so long as $\gamma / (k \beta) < 1$.

\bigskip

\begin{myexam}
\label{exam:dbl_issuance}
In 2020, both BTC and BCH are expected to have mined 18.375e6 total coins and will reduce their coinbase rewards from $c = 12.5$ down to $c = 6.25$ coins in what is called a \emph{halving} event\footnote{\url{https://en.bitcoin.it/wiki/Controlled\_supply}}. At that point, both chains will have completed fraction 0.875 of their total planned issuance of 21e6 coins. Suppose that at the time of the planned halving, BCH decides instead to continue issuing 12.5 coins per block for an additional 4 years, and then resume with the convention of halving the coins every four years after that. This practice ensures that BCH will always award twice as many coins as BTC (that is until both coins eventually cut issuance to 0), i.e. $k=2$. BCH will also emit approximately 2e6 additional coins, which amounts to slightly less that 10\% more than the originally planned issuance. Based on the new issuance, at all times $\beta / \gamma > 18.375e6 / (2e6 + 18.375e6) > 0.9$. Using Equation~\ref{eq:equil_incr}, we find that in the worst-case, this increases the allocation for BCH to $\alpha / (\alpha + 0.55)$ of the total hash rate from $\alpha / (\alpha + 1)$ before issuance was increased. On July 2, 2019, BTC traded for approximately 11,000~USD and BCH traded for roughly 400~USD\footnote{\url{https://coinmarketcap.com}}. This implies that $\alpha \approx 0.036$, which means that the current equilibrium hash rate allocation for BCH is approximately 0.034 of the total, but it would increase to 0.061 of the total after extending its coin issuance, nearly a two-fold increase.
\end{myexam}

\bigskip

\begin{myexam}
\label{exam:perf_issuance}
Expanding on Example~\ref{exam:dbl_issuance}, we can imagine a blockchain $B$ that tunes its issuance in the extreme to achieve a chosen hash rate equilibrium relative to another chain $A$. Suppose that $P_B / P_A = \alpha < 1$, but chain $B$ chooses $c_B = c_A / \alpha$ so that $V_B / V_A = 1$. Theorem~\ref{thm:equil} predicts that as long as $P_B > 0$, $\boldsymbol{w}_e^* = (0.5, 0.5)$. Of course, $c_B > c_A$, so $B$ coins are issued more rapidly than $A$ coins. This means that chain $B$ will have achieved \textbf{parity in security} with chain $A$ at the expense of more rapidly devaluing its coin relative to coin $A$. A natural question is, will the market reward chain $B$ for this increase in security with an increase in market cap? Note that the mechanism proposed here is more complicated than the one proposed in Example~\ref{exam:dbl_issuance} because here we require that the chain $B$ protocol has knowledge of the price ratio $P_B / P_A$. One way that this can be accomplished is for chain $B$ to implement the $\texttt{Oracle}$ contract as described in Section~\ref{sec:oracle}.
\end{myexam}

\subsection{Cost of Loyal Mining}
\label{sec:cost_loyal}

In this section, we attempt to quantify the cost for miners who are loyal to mining a single chain $B$ when there exists an option to mine either chain $A$ or $B$. At times, this choice can be profitable, but at other times, there exists an opportunity cost in the form of higher profits associated with mining on chain $A$. Equation~\ref{eq:lem_dari} gives the hash adjusted reward or HAR vector $\boldsymbol{\pi}$ associated with the allocation vector $\boldsymbol{w} = (xR, y(1-R))$, where $x$ and $y$ are arbitrary positive constants such that $xR + y(1 - R) = 1$. Vector $\boldsymbol{\pi}$ indicates the expected reward per hash performed on each blockchain. Theorem~\ref{thm:equil} identifies a unique choice for $x$ and $y$ that gives an \emph{equilibrium allocation}, $\boldsymbol{w}_e$, which is the only point where the HAR values for each chain are equal. Finally, Theorem~\ref{thm:equil_convg} establishes that when the allocation to chain $B$ is less than the equilibrium; i.e. $w_B < w_{eB}$, the greedy choice is to increase allocation to $B$ and therefore the HAR value is higher on chain $B$ than on chain $A$. Thus, in this regime a miner loyal to chain $B$ will profit. However, once $w_B > w_{eB}$, the opposite is true. 

For miner $m$, define hash rate vector $\boldsymbol{\phi}$ as the allocation for $m$ expressed as a fraction of total hash rate $H$; e.g. $H \phi_X$ gives the hash rate for $m$ on chain $X$. 
The \emph{utility vector} for $m$ is defined as 
\begin{equation}
\label{eq:utility}
\boldsymbol{U}(\boldsymbol{\phi}, \boldsymbol{\pi}) = H \boldsymbol{\phi} \cdot \boldsymbol{\pi},
\end{equation} 
Component $U(\boldsymbol{\phi}, \boldsymbol{\pi})_X$ gives the total fiat value captured by $m$ mining for one second on chain $X$. Note that $\boldsymbol{U}(\boldsymbol{w}, \boldsymbol{\pi})$ gives the aggregate utility for all miners collectively. The \emph{opportunity cost} to $m$ for shifting from allocation $\boldsymbol{\phi}$ to $\boldsymbol{\phi}'$ is given by
\begin{equation}
\label{eq:opp_cost}
\kappa(\boldsymbol{\phi}, \boldsymbol{\pi}; \boldsymbol{\phi}', \boldsymbol{\pi}') = \boldsymbol{U}(\boldsymbol{\phi}, \boldsymbol{\pi}) - \boldsymbol{U}(\boldsymbol{\phi}', \boldsymbol{\pi}').
\end{equation}

\subsubsection{Utility between highly similar blockchains}
\label{sec:util_similar}

In the special case where $T_A = T_B = T$, $\mathcal{S}_A = \mathcal{S}_B$, and $c_A = c_B = c$, the utility for miner $m$ is given by
\[
\boldsymbol{U}(\boldsymbol{\phi}, \boldsymbol{\pi}) = \frac{c(P_A + P_B)}{T} \boldsymbol{\phi} \cdot \left( \frac{1}{x}, \frac{1}{y} \right).
\]
At equilibrium, $x=y=1$ and utility becomes
\[
\boldsymbol{U}(\boldsymbol{\phi})_e = \frac{c \phi (P_A + P_B)}{T},
\]
where $\phi = |\boldsymbol{\phi}|$. Therefore, the opportunity cost to $m$ for mining with allocation vector $\boldsymbol{\phi}'$ is given by
\begin{equation}
\label{eq:opp_cost}
\kappa(\boldsymbol{\phi}', x', y')_e = \frac{c(P_A + P_B)}{T} \left(\phi - \boldsymbol{\phi}' \cdot \left( \frac{1}{x'}, \frac{1}{y'} \right) \right).
\end{equation}

\bigskip

\begin{myexam}
\label{exam:loyal}
Consider blockchains $A$ and $B$ that are similar in the sense of Section~\ref{sec:util_similar}, and assume that $P_B / P_A = \alpha$ such that $\boldsymbol{w}_e = \left( \frac{1}{1+\alpha}, \frac{\alpha}{1+\alpha} \right)$. Assume further that the allocation is initially at equilibrium, i.e. $\boldsymbol{w} = \boldsymbol{w}_e$. Now suppose that a group of miners $m$ loyal to coin $B$, and having total hash rate $\phi = k w_{eB}$, wish to increase chain $B$'s share of the hash rate by a factor $k$, such that $w'_B = k w_{eB}$. 
Since $w'_B$ exceeds the equilibrium allocation to chain $B$, Theorem~\ref{thm:equil_convg} shows that greedy miners will abandon chain $B$ and therefore \textbf{only} loyal miners will mine on chain $B$, i.e. $w'_B = \phi'_B$.
By definition, $R = 1 / (1+\alpha)$, and from Lemma~\ref{lem:dari}, we have that $Rx' = 1 - \frac{k \alpha}{1+\alpha}$ and $y'(1-R) = \frac{k \alpha}{1+\alpha}$. Hence, according to Equation~\ref{eq:opp_cost}, we find the opportunity cost per block to be
\[
\begin{array}{rcl}
T \kappa(\boldsymbol{\phi}', x', y')_e &=& c (P_A + P_A) \left( \phi - \boldsymbol{\phi}' \cdot \left( \frac{1}{x'}, \frac{1}{y'} \right) \right) \\
&=& c (P_A + P_B) \left( \frac{k \alpha}{1+\alpha} - (0, \frac{k \alpha}{1+\alpha}) \cdot \left( \frac{R(1+\alpha)}{1 - \alpha(k-1)}, \frac{(1-R)(1+\alpha)}{k \alpha} \right) \right) \\
&=& c (P_A + P_B) \left( \frac{k \alpha}{1+\alpha} + R - 1 \right) \\
&=& c (P_A + P_B) \left( \frac{(k-1) \alpha}{1+\alpha} \right) \\
&=& c (P_A + P_B) \alpha R (k-1) \\
&=& c (k-1) P_B.
\end{array}
\]
In words, the opportunity cost for miners $m$ to increase the hash rate of chain $B$ by a factor $k$ beyond the equilibrium allocation for 1 block is exactly equal to $k-1$ times the expected coinbase reward from chain $B$.
\end{myexam}

\bigskip

\begin{figure}[t]
\begin{minipage}[c]{0.63\textwidth}
\includegraphics[width=\linewidth]{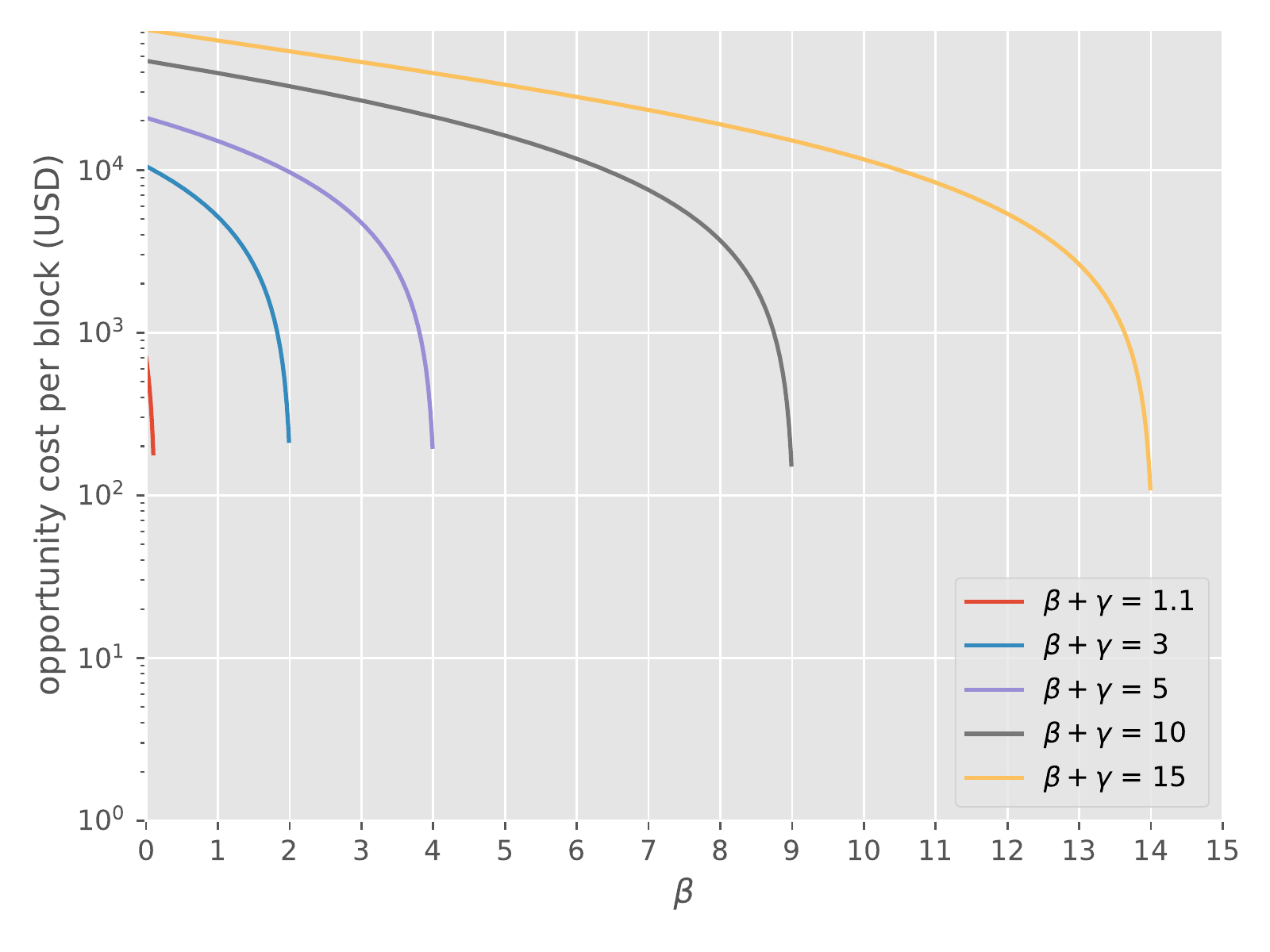}
\end{minipage}
\begin{minipage}[c]{0.35\textwidth}
\caption{Opportunity cost in USD (Equation~\ref{eqn:minority_attack}) to miners from BTC who divert hash rate to BCH in order to cause blockchain reorganization. Attacker diverts multiple $\gamma$ times the equilibrium BCH hash rate ($H_{\text{BCH}}$), where $\gamma > 1$, in order to create a fork of the BCH chain. Attacker also leaves $\beta H_{\text{BCH}}$ to mine on BTC. For hash rate $\beta + \gamma$, opportunity cost is lowest as $\gamma$ approaches 1, but this requires much longer to reorganize the BCH chain. $\beta + \gamma = 15$ corresponds to roughly 55\% of the total hash rate.}
\label{fig:minority_attack}
\end{minipage}\vspace{-2ex}
\end{figure}

\begin{myexam}
\label{exam:minority_attack}
Continuing with Example~\ref{exam:loyal}, suppose that a group of miners $m$ from chain $A$, having aggregate hash weight $(\beta + \gamma) \frac{\alpha}{1 + \alpha}$ where $\gamma > 1$ and $\beta + \gamma < \frac{1}{\alpha}$, conspire to \emph{reorgainze}, i.e. orphan, the last $z$ blocks on chain $B$. They will do this by diverting hash rate $\gamma \frac{\alpha}{1 + \alpha}$ from chain $A$ to a \textbf{fork} of chain $B$. Thus, the existing hash rate on chain $B$, $\frac{\alpha}{1+\alpha}$, will be lost entirely. For simplicity, we assume that both DAAs come to rest immediately (which incurs negligible error when $z$ is large). It follows then that, during the attack, the new hash allocation will be 
\[
\boldsymbol{w}' = (1+\alpha) \left( \frac{1- \alpha \gamma}{1+\alpha},  \frac{\alpha \gamma}{1+\alpha} \right) = \left(1- \alpha \gamma, \alpha \gamma \right).
\] 
with fraction $\frac{\alpha \beta}{1+ \alpha} / (\frac{1}{1 + \alpha} - \frac{\alpha \gamma}{1 + \alpha}) = \frac{\alpha \beta}{1 - \alpha \gamma}$ of $w'_A$ and the entirety of $w'_B$ being controlled by miners $m$. In the parlance of Section~\ref{sec:util_similar}, we have $\boldsymbol{\phi}' = (\alpha \beta, \alpha \gamma)$, $\phi = \alpha (\beta + \gamma)$, $x' = (1+\alpha)(1-\alpha \gamma)$, and $y' = \gamma (1 + \alpha)$. Therefore, using the same reasoning as in Example~\ref{exam:loyal}, the opportunity cost for miners $m$ is equal to
\begin{equation}
\label{eqn:minority_attack}
\begin{array}{rcl}
T \kappa(\boldsymbol{\phi}', x', y')_e &=& c (P_A + P_A) \left( \phi - \boldsymbol{\phi}' \cdot \left( \frac{1}{x'}, \frac{1}{y'} \right) \right) \\
&=& c (P_A + P_B) \left( \alpha(\beta + \gamma) - \left( \alpha \beta, \alpha \gamma \right) \cdot \left( \frac{1}{(1+\alpha) (1 - \alpha \gamma)}, \frac{1}{\gamma(1+\alpha)} \right) \right) \\
&=& c (P_A + P_B) \left( \alpha(\beta + \gamma) - \frac{\alpha \beta}{(1+\alpha) (1 - \alpha \gamma)} - \frac{\alpha}{1+\alpha} \right).
\end{array}
\end{equation}
Figure~\ref{fig:minority_attack} shows the opportunity cost to BTC miners who attempt to carry out a reorganization attack on the BCH chain. We assume here that $P_{\text{BCH}} = 400$, $P_{\text{BTC}} = 11000$, and $\alpha = P_{\text{BCH}} / P_{\text{BTC}} \approx 0.36$ as was the case on July 2, 2019\footnote{https://coinmarketcap.com}. Prior to attack, BCH has fraction $\alpha / (1 + \alpha)$ of the hash rate. The attacker diverts $\gamma \alpha / (1 + \alpha)$, $\gamma > 1$, hash rate to a fork of the BCH chain and leaves fraction $\beta \alpha / (1 + \alpha)$ to mine on BTC. The plot shows that opportunity cost is lowest as $\gamma$ approaches 1, but this also means that the attacker has roughly the same hash rate as honest miners on the other BCH fork. Thus, a cheaper attack will take much longer to reorganize the BCH chain for fixed reorganization depth $z$.
\end{myexam}

\section{Discussion}
\label{sec:discussion}

\subsection{Greedy is obvious, but why cautious?}
\label{sec:greedy_cautious}

Section~\ref{sec:eval} showed empirically that the hash rate allocations among several pairs of the largest blockchain projects by market cap closely follow the equilibrium described in Theorem~\ref{thm:equil}. There are exceptions, where the allocation diverges from equilibrium, but they tend to be short-lived and align closely with events like hard forks. Also in that section, similar results were observed in simulation when the majority of miners follow an $\epsilon$-greedy policy (see Definition~\ref{def:eps_greedy}) for sufficiently small $\epsilon$. This suggests that much of miner behavior can be explained by a preference for improving immediate reward, but not to an extreme. Specifically, the simulation also showed that for a choice of $\epsilon$ that is too large, allocations oscillate wildly, a phenomenon not typically observed in practice. 

So if mining on a particular chain is currently more profitable than mining on another, why don't miners fully allocate to that chain, i.e. follow the extreme greedy policy? Prior works discussed previously in Section~\ref{sec:related} provide possible explanations. Chatzigiannis et al.~\cite{Chatzigiannis:2019} suggested that miners incur less risk in the form of variance in block reward by mining simultaneously in a mixture of pools and across blockchains. Therefore, there exists incentive to mine at least partially on the less profitable chain in order to enjoy lower variance in payout. Most chains also impose a \emph{cool-down period}\footnote{https://bitcoin.org/en/blockchain-guide\#transaction-data} for newly awarded coins during which they cannot be spent.  Bissias et al.~\cite{Bissias:2018} argued that this imparts risk to the miner in the form of price volatility during the cool-down period. They showed that miners can minimize risk by allocating their hash rate to a mixed \emph{portfolio} of blockchains. Thus, again, the extreme greedy policy may be inferior to a mixed strategy that reduces miner risk

\subsection{Implications for Minority Hash Rate Chains}

A major conclusion from Kwon et al.~\cite{Kwon:2019} is that minority hash rate blockchains such as Bitcoin Cash (BCH) might be doomed to fail due to a lack of genuine miner interest. Their reasoning is that, if there exists a loyal miner base devoted to BCH that exceeds the equilibrium allocation, then no profit seeking miners will also mine BCH. Thus, the loyal miners will be alone in propping up the blockchain. While we do not dispute the possibility of this scenario, it is also not clearly a likely outcome. First, Theorem~\ref{thm:equil_convg} proves that for greedy but cautious miners, there exists a tendency to move toward equilibrium. In Section~\ref{sec:eval}, we demonstrated empirically that this tendency is typically manifested in the real world. And at equilibrium, there exists no \emph{preference} to mine one chain over the other. Thus, there is typically no need for loyal mining to maintain hash rate. Second, Example~\ref{exam:loyal} shows that loyal mining beyond the equilibrium point incurs a cost linear in the value of the coinbase reward of the minority chain. Therefore, loyal miners who are actively propping up the hash rate of a blockchain are financially disincentivized from continuing this practice over the long-term, which will also tend to move hash rate allocation back to equilibrium.

The examples in Section~\ref{sec:security} illustrate that the allocation equilibrium point itself is quite fluid, depending mainly on the value of the coinbase reward. As discussed in Section~\ref{sec:related}, this is a concept that was first suggested in abstract by Spiegelman et al.~\cite{Spiegelman:2018}, and we have extended it by quantifying the change in equilibrium given a specific change in coinbase reward. The implication of these results is that minority hash rate blockchains can significantly increase their security relative to the majority hash rate blockchain by simply adjusting their coinbase reward. We further demonstrated that this increase in security can be accomplished for BCH without significantly devaluing the currency.

Finally, for minority hash rate blockchains, there exists a danger that miners from the majority hash rate chain will force a long reorganization of $z$ previously confirmed blocks. In Example~\ref{exam:minority_attack}, we derived an expression for the opportunity cost to attackers from the heavier weight blockchain. For BCH versus BTC (Figure~\ref{fig:minority_attack}), every reorganization costs at least 100~USD per block. However, the lowest cost attacks only allow the attacker to match the hash rate on BCH. This means that the reorganization of many blocks will likely take a long time since it is required that the attacker mine $n+z+1$ blocks in the time the honest miners mine $n$. Cost rises exponentially as the attacker increases hash rate beyond honest BCH miners. For example, if the attackers double the honest hash rate on BCH, then the opportunity cost jumps to at least 3,000~USD per block for any set of attackers with less that 50\% of the total BTC + BCH hash rate.

\section{Conclusion}

In this paper, we have shown formally that a singular hash rate equilibrium arises for miners who split their hash rate among two blockchains assuming that the miners are both greedy and cautious. If they become overly greedy, then their hash rate will oscillate in the extreme between the two chains. Assuming an efficient market for buying and selling hash rate, the results also hold between two blockchains with different PoW algorithms, and even between PoW and PoS blockchains where hash rate is replaced by the opportunity cost associated with locking up stake in the PoS system. We demonstrated these theoretical results empirically using historical data from real world blockchains and data from a block mining simulator. Finally, we presented several applications including a trustless price-ratio oracle, enhanced security for minority hash rate blockchains, and quantification of loyal mining costs.

\section{Acknowledgements}

We would like to thank David Jensen and Akanksha Atrey for many thought provoking discussions, which helped us to focus our investigation. We would also like to thank Rainer B{\"o}hme for his insights in the discussions we had with him.

\urlstyle{sf}
\pagestyle{plain}

{\footnotesize \bibliographystyle{acm}
\bibliography{references}}

\appendix

\end{document}